%% file: chapternu.tex
\documentclass[11pt,fleqn]{article}

\usepackage{graphicx}    

\newcommand{\nuh}[1]{\nu_{#1}^{\mbox{{\scriptsize Heavy}} }}
\newcommand{\nul}[1]{\nu_{#1}^{\mbox{{\scriptsize Light}} }}
\newcommand{\scr}[1]{{\cal {#1}}_\nu}
\newcommand{\scrl}{{\cal L}_{m_\nu}}
\newcommand{\Eq}[1]{Eq.~(\ref{eq#1})}
\newcommand{\beq}{\begin{equation}}
\newcommand{\eeq}{\end{equation}}

\begin{document}

\author{Boris Kayser \\ 
{\small Theoretical Physics Department }\\
{\small Fermi National Accelerator Laboratory, P.O. Box 500, Batavia, IL 60510} }
\title{Neutrino Mass, Mixing, and Flavor Change
\thanks{To appear in {\it Neutrino Mass},  eds. G. Altarelli and K. Winter (Springer Tracts in Modern Physics).} }
\maketitle

\begin{abstract}
The theoretical basics of neutrino mass and mixing are reviewed. Dirac and Majorana masses are explained, and added together to produce the see-saw picture of the lightness of neutrinos. This picture predicts that neutrinos are Majorana particles. The character, and an apparent paradox, of Majorana neutrinos are examined. The physics of neutrino flavor change (oscillation), {\it in vacuo} and in matter, is reviewed.
\end{abstract}

\newpage

\section{Neutrino masses and mixing, and the see-saw}
\label{s1}

The evidence that neutrinos change from one flavor to another is compelling \cite{r1}. Barring exotic possibilities, neutrino flavor change implies neutrino mass and mixing. Thus, neutrinos almost certainly have nonzero masses and mix.

That neutrinos have masses means that there is a spectrum of three or more neutrino mass eigenstates, $\nu_1, \nu_2, \nu_3, \ldots$. That neutrinos mix means that the neutrino state coupled by the charged-current weak interaction to the $W$ boson and a specific charged lepton (such as the electron) is none of the neutrino mass eigenstates, but rather is a mixture of them. 
Consider, for example, the leptonic $W$ decay $W^+ \rightarrow \ell_\alpha^+ + \nu_\alpha$, yielding the specific charged lepton $\ell_\alpha$. Here, the ``flavor'' $\alpha$ of the lepton can be $e, ~\mu,$ or $\tau$, and $\ell_e$ is the electron, $\ell_\mu$ the muon, and $\ell_\tau$ the $\tau$. In the $W$ decay, the produced neutrino state $|\nu_\alpha \rangle$, referred to as the neutrino of flavor $\alpha$, is the superposition
\beq
|\nu_\alpha \rangle = \sum_i U^*_{\alpha i} |\nu_i \rangle 
\label{eq1}
\eeq
of the mass eigenstates $|\nu_i \rangle$. Here, $U$ is a matrix known as the leptonic mixing matrix \cite{r2}.

\begin{figure}[hbt]
\begin{center}
\includegraphics[height=4cm]{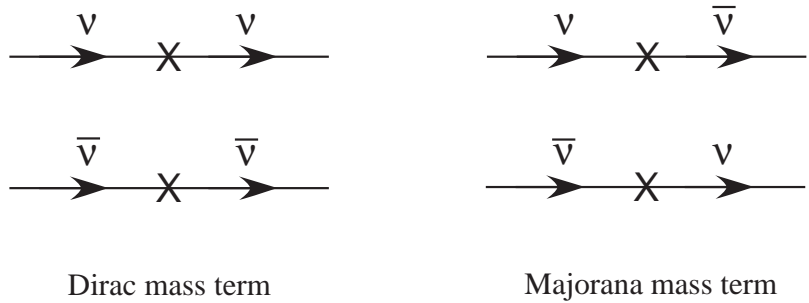}
\end{center}
\caption{The effects of Dirac and Majorana mass terms. The action of the mass terms is represented by the symbol X.}
\label{f1}
\end{figure}

Through our studies of neutrinos, we hope to eventually discover what physics lies behind their masses and mixing \cite{r3}. This underlying physics may contain neutrino mass terms of two different kinds: Dirac and Majorana. As depicted in Fig.~\ref{f1}, a Dirac mass term turns a neutrino into a neutrino, or an antineutrino into an antineutrino, while a Majorana mass term converts a neutrino into an antineutrino, or vice versa. 
Thus, Dirac mass terms conserve the lepton number $L$ that distinguishes leptons from antileptons, while Majorana mass terms do not. The quantum number $L$ is also conserved by the Standard Model (SM) couplings of neutrinos to other particles. 
Thus, if we assume that the interactions between neutrinos and other particles are well described by these SM couplings---a very plausible assumption in view of the great success of the SM---then any $L$ nonconservation that we might oberve in neutrino experiments would have to arise from Majorana mass terms, not from interactions.

A Dirac mass term $D$ may be constructed out of a chirally left-handed neutrino field $\nu_L^0$, and a chirally right-handed one $\nu_R^0$ \cite{r4}:
\beq
{\cal L}_D = -m_D \overline{\nu_R^0} \nu_L^0 + \mbox{ h. c.}~~.
\label{eq2}
\eeq
A Majorana mass term may be constructed out of $\nu_L^0$ alone, in which case we have the ``left-handed Majorana mass''
\beq
{\cal L}_{m_L} = -\frac{m_L}{2} \overline{(\nu_L^0)^c} \nu_L^0 + \mbox{ h. c.}~~,
\label{eq3}
\eeq
or out of $\nu_R^0$ alone, in which case we have the ``right-handed Majorana mass''
\beq
{\cal L}_{m_R} = -\frac{m_R}{2} \overline{(\nu_R^0)^c} \nu_R^0 + \mbox{ h. c.}~~.
\label{eq4}
\eeq
In these expressions, $m_D, m_L$, and $m_R$ are mass parameters, and for any field $\psi$, $\psi^c$ is the corresponding charge-conjugate field. In terms of $\psi,~ \psi^c = C\bar{\psi}^T$, where $C$ is the charge conjugation matrix, and $T$ denotes transposition.

(In writing both mass and interaction terms, we use the superscript ``zero'' to denote underlying fields out of which a model is constructed. Fields without a superscript zero correspond to physical particles of definite mass.)

Table \ref{t1} indicates the effects of the various fields in the mass terms on neutrinos and antineutrinos. From this table, we see that each type of mass term does indeed induce the transitions ascribed to it in Fig.~\ref{f1}.

\begin{table}[hbt]
\begin{center}
\caption{The effects of the fields. ``A'' signifies that the given particle is annihilated by the field, and ``C'' that the particle is created.}
\label{t1}    
\begin{tabular}{lcc}
\noalign{\smallskip}\hline\noalign{\smallskip}
Field                     & Effect on $\nu$ & Effect on $\bar{\nu}$  \\
\noalign{\smallskip}\hline\noalign{\smallskip}
$\nu_{L,R}^0$        &             A             &        C  \\ \noalign{\smallskip} \\
$\overline{\nu_{L,R}^0}$  & C             &         A  \\ \noalign{\smallskip}\\
$(\nu_{L,R}^0)^c $ &             C             &        A   \\ \noalign{\smallskip}\\
$\overline{(\nu_{L,R}^0)^c}$ &   A      &         C   \\ 
\noalign{\smallskip}\hline
\end{tabular}
\end{center}
\end{table}

An electrically charged fermion such as a quark cannot have a Majorana mass term, because such a term would convert it into an antiquark, in violation of electric charge conservation. However, for the electrically neutral neutrinos, Majorana mass terms are not only allowed but rather likely, given that the neutrinos are now known to be particles with mass \cite{r5}. 
To see why the Majorana mass terms are likely, suppose first that some neutrino is described by the SM. In the original version of that model, the neutrino would be massless. It would be described by the left-handed (LH) field that we have called $\nu_L^0$, and the model would contain no right-handed (RH) counterpart to $\nu_L^0$. 
Let us suppose that we now try to extend the SM to accommodate a nonzero mass for this neutrino in the same way that the SM already accommodates nonzero masses for the quarks and charged leptons. The latter masses, of course, are all of Dirac type, and arise from Yukawa couplings of the form
\beq
-f_q \varphi \overline{(q_L^0)} q_R^0 + \mbox{ h. c.}~~.
\label{eq5}
\eeq
Here, $q^0$ is some quark, $\varphi$ is the neutral Higgs field, and $f_q$ is a coupling constant. When $\varphi$ develops a vacuum expectation value $\langle \varphi \rangle_0$, the coupling of \Eq{5} yields a term
\beq
-f_q \langle \varphi \rangle_0 \overline{(q_L^0)} q_R^0 + \mbox{ h. c.} ~~ .
\label{eq6}
\eeq
This is a Dirac mass term of the form of Eq.~(\ref{eq2}) for the quark $q^0$ with $-f_q \langle \varphi \rangle_0$ the mass. Extending the SM to include a mass for our neutrino that parallels the masses of the quarks is a simple matter of adding to the model a RH neutrino field $\nu_R^0$ and a Yukawa coupling $-f_\nu \varphi \overline{(\nu_L^0)} \nu_R^0 +$ h. c., with $f_\nu$ a suitable coupling constant. When $\varphi$ develops its vacuum expectation value, this coupling will yield the Dirac mass term
\beq
{\cal L}_D = -f_\nu \langle \varphi \rangle_0 \overline{(\nu_L^0)} \nu_R^0 +\mbox{ h. c.}
\label{eq7}
\eeq
for this neutrino. This term imparts to the neutrino a mass $m_\nu = f_\nu \langle \varphi \rangle_0$. Now suppose, for example, that we would like $m_\nu$ to be of order 0.05 eV, the neutrino mass scale suggested by the observed atmospheric neutrino oscillations. Since $\langle \varphi \rangle_0 =$ 174 GeV, the coupling $f_\nu$ must then be of order 10$^{-13}$. 
Such an infinitesimal coupling constant may not be out of the question, but it certainly strikes one as unlikely to be the ultimate explanation of neutrino mass.

In addition, to generate a Dirac mass for the neutrino, we were obliged to introduce the RH neutrino field $\nu_R^0$. In the SM, right-handed fermion fields are weak-isospin singlets. Hence, so are their charge-conjugates. Thus, once the field $\nu_R^0$ exists, there is nothing in the SM to prevent the occurrence of a right-handed Majorana mass term like that in \Eq{4}: 
Such a term violates neither the conservation of weak isospin nor that of electric charge. Consequently, if nature contains a Dirac neutrino mass term, then it is highly likely that she contains a Majorana mass term as well. And, needless to say, if nature does not contain a Dirac neutrino mass term, then she certainly contains a Majorana mass term, which would then be the only source of neutrino mass.

Suppose that a neutrino has a Dirac mass, as the quarks and charged leptons do, and also a right-handed Majorana mass like that in \Eq{4}, as suggested by the previous argument. Then its total mass term $\scrl$ is
\begin{eqnarray}
\scrl &  = & -m_D \overline{\nu_R^0} \nu_L^0 - \frac{m_R}{2} \overline{(\nu_R^0)^c}\, \nu_R^0 + \mbox{ h. c.}  \nonumber \\
& = &  - \, \frac{1}{2} \,[\overline{(\nu_L^0)^c}, \overline{\nu_R^0} ]
	\left[  \begin{array}{cc}
		 0 & m_D   \\ m_D & m_R
	\end{array} \right] 
	\left[  \begin{array}{c}
		 \nu_L^0  \\  (\nu_R^0)^c
	 \end{array}  \right] + \mbox{ h. c.}
\label{eq7a}
\end{eqnarray}
Here, we have used the identity $\overline{(\nu_L^0)^c}  m_D (\nu_R^0)^c = \overline{\nu_R^0}  m_D \nu_L^0$. The matrix
\beq
\scr{M} = \left[  \begin{array}{cc}
		 0 & m_D \\ m_D & m_R
						 \end{array} \right] 
\label{eq8}
\eeq	
appearing in $\scrl$ is referred to as the neutrino mass matrix.

It is natural to suppose that the Dirac mass $m_D$ of our neutrino is of the same order of magnitude as the Dirac masses of the quarks and charged leptons, since in the SM all of these Dirac masses arise from couplings to the same Higgs field. Of course, the Dirac masses of the quarks and charged leptons are their total masses, so we expect $m_D$ to be of the same order of magnitude as a typical quark or charged lepton 	mass. Furthermore, since nothing in the SM requires the right-handed Majorana mass $m_R$ to be small, we expect that this mass is {\em large}: $m_R\gg m_D$.

The mass matrix $\scr{M}$ can be diagonalized by the transformation
\beq
Z^T \scr{M} Z = \scr{D}~~,
\label{eq9}
\eeq
where
\beq
\scr{D} = \left[  \begin{array}{cc}
							m_1 & 0 \\ 0 & m_2
							\end{array}  \right]
\label{eq10}
\eeq
is a diagonal matrix whose diagonal elements are the positive-definite eigenvalues of $\scr{M}$ \cite{r6}, $Z$ is a unitary matrix, and $T$ denotes transposition. To first order in the small parameter $\rho \equiv m_D/m_R$,
\beq
Z =            \left[  \begin{array}{cc}
							1 & \rho \\ -\rho & 1
							\end{array}  \right]  \;
				\left[  \begin{array}{cc}
							i & 0 \\ 0 & 1
							\end{array}  \right]~~.
\label{eq11}
\eeq
Using this $Z$ in \Eq{9}, one finds that to order $\rho^2$,
\beq
\scr{D} = \left[  \begin{array}{cc}
							m_D^2 / m_R & 0 \\ 0 & m_R
							\end{array}  \right]~~.
\label{eq12}
\eeq	
Thus, the mass eigenvalues are $m_1 \simeq  m_D^2 / m_R$ and $m_2 \simeq m_R$.

To recast $\scrl$ in terms of mass eigenfields, we define the two-component column vector $\nu_L$ by
\beq
\nu_L  \equiv  Z^{-1} \left[  \begin{array}{c}
						 \nu_L^0  \\  (\nu_R^0)^c
									 \end{array}  \right] ~~.
\label{eq13}
\eeq
(The column-vector field $\nu_L$ is chirally left-handed, since the charge conjugate of a field with a given chirality always has the opposite chirality.) We then define the two-component field $\nu$, with components $\nu_1$ and $\nu_2$, by
\beq
\nu \equiv \nu_L + (\nu_L)^c \equiv 
		\left[  \begin{array}{c}
		\nu_1  \\  \nu_2
		\end{array}  \right] ~~ .
\label{eq14}
\eeq
Using the fact that scalar covariant combinations of fermion fields can connect only fields of opposite chirality, it is easy to show that the $\scrl$ of \Eq{7a} may be rewritten as
\beq
\scrl = -\sum_{i=1}^2 \frac{m_i}{2} \overline{\nu_i}\nu_i ~~ .
\label{eq15}
\eeq
We recognize the $i$'th term of this expression as the usual mass term for a neutrino $\nu_i$. The mass of that neutrino appears to be $m_i / 2$, but we shall see shortly that it is actually $m_i$.

From the definition of \Eq{14}, we see that $\nu_i = \nu_{Li} + \nu_{Li}^c$ goes into itself under charge conjugation. A neutrino whose field has this property is identical to its antiparticle \cite{r7}, and is known as a Majorana neutrino. Thus, the eigenstates of the combined Dirac-Majorana mass term $\scrl$ of \Eq{7a} are Majorana neutrinos.

Fermions that are distinct from their antiparticles are known as Dirac particles. The mass term for a Dirac fermion $f$ of mass $m_f$ is $-m_f \bar{f} f$. But the mass term for a Majorana neutrino $\nu$ of mass $m_\nu$ is -(1/2) $m_\nu \bar{\nu} \nu$. 
To see why there is this extra factor of 1/2, we note that if $\nu$ has a mass term $-k\bar{\nu}\nu$ in the Lagrangian density (with $k$ some constant), then the mass of $\nu$ is
\beq
\langle \nu \,\mbox{ at rest } | \int d^3x \,k\bar{\nu}\nu |\; \nu \mbox{ at rest} \rangle ~~ .
\label{eq16}
\eeq
If $\nu$ is a Majorana particle, this matrix element is twice as large as it would be if $\nu$ were a Dirac particle. To see why, suppose first that $\nu$ is a Dirac particle ($\bar{\nu} \neq \nu$). Then the field $\nu$ can absorb a neutrino or create an antineutrino. The field $\bar{\nu}$ can absorb an antineutrino or create a neutrino. 
Thus, in the matrix element (\ref{eq16}), it is the field $\nu$ that absorbs the initial neutrino, and the field $\bar{\nu}$ that creates the final one. Now suppose that $\nu$ is a Majorana particle. Then  the fields $\nu$ and $\bar{\nu}$ still do just what they did in the Dirac case, except that now there is no difference between the ``antineutrino'' and the neutrino. 
The field $\nu$ can either absorb or create this neutrino, and so can the field $\bar{\nu}$. Thus the matrix element (\ref{eq16}) has two terms: In the first, the field $\nu$ absorbs the initial neutrino and the field $\bar{\nu}$ creates the final one. In the second, the field $\bar{\nu}$ absorbs the initial neutrino and the field $\nu$ creates the final one. It is straightforward to show that these two terms are equal, and that each of them is equal to the single term present in the Dirac case. Hence, for a given $k$, the matrix element (\ref{eq16}) is twice as big in the Majorana case as in the Dirac one. As is well known, in the Dirac case it is just equal to $k$, so that in the Majorana case $k$ is half the mass.

With $m_D$ of the order of a typical quark or charged lepton mass, and $m_R \gg m_D$, the mass of $\nu_1$,
\beq
m_1 \cong m_D^2 / m_R ~~ ,
\label{eq17}
\eeq
can be very small. Thus, if we identify $\nu_1$ as one of the light neutrinos, we have an elegant explanation of why it is so light. This explanation, in which physical neutrino masses are small because the RH Majorana mass $m_R$ is large, is known as the see-saw mechanism, and \Eq{17} is referred to as the see-saw relation \cite{r8}. 
The mass $m_R$ is assumed to reflect some high mass scale where new physics responsible for neutrino mass resides. Interestingly, if $m_R$ is just a bit below the grand unification scale---say $m_R \sim 10^{15}$ GeV---and $m_D \sim m_{\mbox{\scriptsize top}} \simeq 175$ GeV, then from \Eq{17} $m_1 \sim 3 \times 10^{-2}$ eV. This is right in the range of neutrino mass suggested by the experiments on atmospheric neutrino oscillation \cite{r9}.

The reader will have noticed that under our assumptions about $m_D$ and $m_R$, the mass of $\nu_2$,
\beq
m_2  \cong  m_R ~~ ,
\label{eq18}
\eeq
is far from small. The eigenstate $\nu_2$ cannot be one of the light neutrinos, but is a hypothetical very heavy neutral lepton. Such neutral leptons figure prominently in attempts to explain the baryon-antibaryon asymmetry of the universe in terms of leptogenesis.

The see-saw mechanism, based as it is on the $\scrl$ of \Eq{7a}, predicts that the light neutrinos such as $\nu_1$, as well as the hypothetical heavy neutral leptons such as $\nu_2$, are Majorana particles. The light neutrino aspect of this prediction is one of the factors driving a major effort \cite{r10} to look for neutrinoless double beta decay. 
This is the L-violating reaction Nucl $\rightarrow$ Nucl$^\prime$ + 2e$^-$, in which one nucleus decays to another plus two electrons. Observation of this reaction at any nonzero level would show that the light neutrinos are indeed Majorana particles \cite{r11}.

So far, we have analyzed the simplified case in which there is only one light neutrino and one heavy neutral lepton. In the real world, there are three leptonic generations, with a light neutrino in each one, and the particles in different generations mix. It is quite easy to extend our analysis to accommodate this situation \cite{r12}.

In the SM, there are left-handed weak-eigenstate charged leptons $\ell_{L\alpha}^0$, with $\alpha = e, \; \mu$, and $\tau$. Each $\ell_{L\alpha}^0$ couples to a LH weak-eigenstate neutrino $\nu_{L\alpha}^0$ via the charged-current weak interaction
\beq
{\cal L}_W = -\frac{g}{\sqrt{2}}W_\rho^- \sum_{\alpha=e,\mu,\tau} \overline{\ell_{L\alpha}^0} \gamma^\rho \nu_{L\alpha}^0 + \mbox{ h. c.} ~~ .
\label{eq19}
\eeq
Here, $W$ is the charged weak boson, and $g$ is the semiweak coupling constant. To allow for neutrino masses, one adds to the model RH fields $\nu_{R\alpha}^0$, where $\alpha = e,\;\mu$, or $\tau$. Then, in analogy with \Eq{7a}, one introduces the neutrino mass term
\beq
\scrl  = - \frac{1}{2} \,[\overline{(\nu_L^0)^c}, \, \overline{\nu_R^0 }]
	\left[  \begin{array}{cc}
		 0 & m_D^T   \\ m_D & m_R
	\end{array} \right] 
	\left[  \begin{array}{c}
		 \nu_L^0  \\  (\nu_R^0)^c
	 \end{array}  \right] + \mbox{ h. c.} ~~ .
\label{eq20}
\eeq
Here, $\nu_L^0$ is the column vector
\beq
\nu_L^0 \equiv \left[ \begin{array}{c}
		\nu_{Le}^0  \\ \nu_{L\mu}^0  \\ \nu_{L\tau}^0 
		                   \end{array}  \right] ~~ ,
 \label{eq21}
 \eeq
and similarly for $\nu_R^0$. The quantities $m_D$ and $m_R$ are now 3x3 matrices. In writing \Eq{20}, we have used the fact that, for a given $\alpha$ and $\beta$, $\overline{(\nu_{L\alpha}^0)^c} (m_D^T)_{\alpha\beta} (\nu_{R\beta}^0)^c$ $= \overline{(\nu_{R\beta}^0)}(m_D)_{\beta\alpha} (\nu_{L\alpha}^0)$. 
Thus, once one sums on $\alpha$ and $\beta$, the contributions of the submatrices $m_D$ and $m_D^T$ to $\scrl$ are identical, and add up to conventional Dirac mass terms without the factor of 1/2 at the front of $\scrl$. Since $\overline{(\nu_{R\alpha}^0)} (\nu_{R\beta}^0)^c = \overline{(\nu_{R\beta}^0)} (\nu_{R\alpha}^0)^c$, the matrix $m_R$ may be taken to be symmetric. Thus, the 6x6 mass matrix
\beq
\scr{M} = \left[  \begin{array}{cc}
		 0 & m_D^T \\ m_D & m_R
						 \end{array} \right] 
\label{eq22}
\eeq	
is symmetric. Such a matrix may be diagonalized by the transformation of \Eq{9}, but with $Z$ now a 6x6 unitary matrix and $\scr{D}$ a 6x6 diagonal matrix whose diagonal elements $m_i,\; i=1,\dots,6$, are the positive-definite eigenvalues of $\scr{M}$, \Eq{22}.

To re-express $\scrl$ in terms of mass-eigenstate neutrinos, one introduces the column vector $\nu_L$ via \Eq{13} as before. Of course, in that relation $\nu_L^0$ and $(\nu_R^0)^c$ now each have three components, $Z$ is 6x6, and $\nu_L$ has six components. One then introduces the field $\nu$ via a six-component version of \Eq{14}:
\beq
\nu \equiv \nu_L + (\nu_L)^c \equiv 
		\left[  \begin{array}{c}
		\nu_1  \\  \nu_2 \\ \vdots \\ \nu_6
		\end{array}  \right] ~~ .
\label{eq23}
\eeq
It is then easily shown, as before, that the mass term $\scrl$ of \Eq{20} may be rewritten as
\beq
\scrl = -\sum_{i=1}^6 \frac{m_i}{2} \overline{\nu_i} \nu_i ~~ .
\label{eq24}
\eeq
Thus, the $\nu_i$ are the neutrinos of definite mass, the mass of $\nu_i$ being $m_i$. From \Eq{23}, we see that each $\nu_i$ is a Majorana neutrino.

To complete the treatment of the leptonic sector, one introduces for the charged leptons a (Dirac) mass term ${\cal L}_{m_\ell}$ given by \cite{r13}
\beq
{\cal L}_{m_\ell} = -\overline{\ell^0_R} {\cal M}_\ell \ell^0_L + \mbox{ h. c.} ~~ .
\label{eq25}
\eeq
Here,
\beq
\ell^0_L \equiv \left[ \begin{array}{c}
		 \ell^0_{Le} \\ \ell^0_{L\mu} \\ \ell^0_{L\tau}
		\end{array}  \right]
\label{eq26}
\eeq
is a column vector whose $\alpha$'th component is the LH weak-eigenstate charged lepton field $\ell^0_{L\alpha}$. The quantity $\ell^0_R$ is an analogous column vector whose $\alpha$'th component is the RH weak-isospin singlet charged lepton field $\ell^0_{R\alpha}$. Finally, ${\cal M}_\ell$ is the 3x3 charged lepton mass matrix. This matrix may be diagnonalized by the transformation \cite{r13}
\beq
A^\dagger_R {\cal M}_\ell A_L = {\cal D}_\ell ~~ ,
\label{eq27}
\eeq
where $A_{L,R}$ are two distinct 3x3 unitary matrices, and
\beq
{\cal D}_\ell = \left[  \begin{array}{ccc}
		m_e &    0        &   0    \\
		   0   & m_\mu &   0    \\
		   0   &    0        & m_\tau \\
		\end{array}  \right]
\label{eq28}
\eeq
is the diagonal matrix whose diagonal elements are the charged lepton masses.

If one defines the three-component column vectors $\ell_{L,R}$ via
\beq
\ell^0_{L,R} = A_{L,R}\, \ell_{L,R} ~~ ,
\label{eq29}
\eeq
and then introduces the vector
\beq
\ell \equiv \ell_L + \ell_R ~~ ,
\label {eq30}
\eeq
one quickly finds that
\beq
{\cal L}_{m_\ell} = -\overline{\ell}\, {\cal D}_\ell \,\ell 
  = - \sum_{\alpha=e,\mu,\tau} \overline{\ell_\alpha} m_\alpha \ell_\alpha ~~ .
\label{eq31}
\eeq
Thus, the components $\ell_\alpha$ of the vector $\ell$ are the charged leptons of definite mass: $e, \;\mu$, and $\tau$. 

To recast the SM weak interaction ${\cal L}_W$, \Eq{19}, in terms of mass eigenstates, it is convenient to write the 6x6 matrix $Z$ in the form
\beq
Z = \left[ \begin{array}{cc}
	V & Y \\ X & W
	\end{array} \right] ~~ ,
\label{eq32}
\eeq
in which $V,\; W,\; X$, and $Y$ are 3x3 submatrices. If the Dirac mass matrix $m_D$ is much smaller than the Majorana mass matrix $m_R$, then $X$ and $Y$ are much smaller than $V$ and $W$ for the same reason as the off-diagonal elements of the 2x2 version of $Z$, \Eq{11}, are small. 
Similarly, from \Eq{12} we may conclude that in the three-generation, six-neutrino, case, the first three neutrinos, $\nu_{1,2,3}$, are light, but the second three, $\nu_{4,5,6}$ are very heavy. To emphasize this, we shall call the first three neutrinos $\nul{1,2,3}$ and the second three $\nuh{1,2,3}$. From experimental searches for heavy neutral leptons, we know that there are none with masses below 80 GeV \cite{r14}. 
Thus, in neutrino experiments at energies less that this (and even at much higher energies if the heavy neutrinos are at the TeV or even the grand unification scale), it is only the light neutrinos that play a significant role. Now, from the 6x6 analogue of \Eq{13} and from \Eq{32}
\beq
\nu^0_{L\alpha}  =  \sum_{i=1}^3 [V_{\alpha i} \nul{Li} + Y_{\alpha i} \nuh{Li}] 
\cong \sum_{i=1}^3 V_{\alpha i} \nul{Li} ~~ ,
 \label{eq33}
\eeq
where in the second step we have used $Y\ll V$. From Eqs.~(\ref{eq33}) and (\ref{eq29}), we may rewrite the weak interaction, \Eq{19} as
\beq
{\cal L}_W \cong -\frac{g}{\sqrt{2}} W_\rho^- \overline{\ell_L} \gamma^\rho U \nul{L} - \frac{g}{\sqrt{2}} W_\rho^+\overline{\nul{L}} \gamma^\rho U^\dagger \ell_L ~~ .
\label{eq34}
\eeq
Here,
\beq
\nul{L}  \equiv  \left[ \begin{array}{l}
				\nul{L1} \\ \nul{L2} \\ \nul{L3} 
				\end{array}  \right]
\label{eq35}
\eeq
is a column vector whose $i$'th component is the left-handed projection of the field of the $i$'th light neutrino mass eigenstate, and
\beq
U \equiv A^\dagger_L V
\label{eq36}
\eeq
is the ``leptonic mixing matrix.'' This is the same matrix as the one called $U$ in \Eq{1}. However, we are now assuming that there are only 3 light neutrinos, so that $U$ is 3x3, and we are relating $U$ to the matrices $A_L$ and $V$ that take part in the diagonalization of the underlying charged lepton and neutrino mass matrices.

\Eq{34} expresses the charged-current weak interaction in terms of charged leptons and neutrinos of definite mass. Since the matrix $Z$ is unitary, and $X$ and $Y$ are much smaller than $V$ and $W$, the matrix $V$ is to a good approximation unitary all by itself. From the unitarity of $A_L$ and \Eq{36}, this means that the leptonic mixing matrix $U$ is approximately unitary as well.

It is not hard to count the number of independent paramenters necessary to fully determine $U$ \cite{r15}. This matrix has 9 entries, each of which may have a real and an imaginary part, for a total of 18 parameters. On these parameters, unitarity imposes 9 constraints: First of all, each of the three columns of $U$ must be a vector of unit length. 
Secondly, each pair of columns must be orthogonal to each other. There are three pairs, and the orthogonality condition for each pair has both a real and an imaginary part, for a total of 6 constraints. With the 9 unitarity constraints taken into account, 9 parameters are left.

With the charged lepton and neutrino indices indicated explicitly, \Eq{34} for the weak interaction reads
\beq
{\cal L}_W =  -\frac{g}{\sqrt{2}} W_\rho^- 
\sum_{\stackrel{\scriptstyle \alpha=e,\mu,\tau} {i=1,2,3} }
\overline{\ell_{L\alpha}} \gamma^\rho U_{\alpha i} \nul{Li} + \mbox{ h. c.} ~~ .
\label{eq37}
\eeq
From \Eq{37}, we see that, apart from the overall strength factor $g/\sqrt{2}, \; U_{\alpha i}$ is essentially just the amplitude $\langle \ell_\alpha^- | H_W | \nul{i}\rangle$ for the transition $\nul{i} \rightarrow \ell_\alpha^-$ via emission or absorption of a $W$ boson, caused by action of the weak Hamiltonian $H_W$ corresponding to ${\cal L}_W$. 
Now we are always free to redefine what we mean by the state $\langle \ell_\alpha^- |$ by multiplying it by a phase factor: $\langle \ell_\alpha^- | \rightarrow e^{i\varphi_\alpha}\langle\ell_\alpha^- |$. Obviously, this phase re-definition causes the $U_{\alpha i}$ for all $i$ to undergo the change $ U_{\alpha i} \rightarrow e^{i\varphi_\alpha} U_{\alpha i}$. Thus, phase re-definition of the 3 charged leptons can be used to remove 3 phase parameters from $U$, leaving a matrix that contains 9 - 3 = 6 parameters. One might think that additional phase parameters could be removed by phase re-definition of the neutrinos. 
If the neutrinos are Dirac particles, this is true. But if, as we are assuming, they are Majorana particles, then one can show that phases removed from $U$ by phase redefining the neutrinos simply show up somewhere else, and still have the same physical effects as they do when they are located in $U$ \cite{r15}. Thus, we shall leave them in $U$, which consequently retains 6 parameters. 
These may be chosen to be mixing angles, which would be present even if $U$ were real, and complex phase factors. To see how many of the 6 parameters are mixing angles, and how many are complex phase factors, we assume for a moment that the latter are turned off (set to unity), so that $U$ is real. It then contains 9 real entries. On these entries, unitarity imposes 6 constraints: 
Each column of $U$ must be a vector of unit length and each pair of columns must be orthogonal. Thus, when the complex phase factors are turned off, $U$ contains 9 - 6 = 3 independent parameters---the mixing angles. Since the complex $U$ with the complex phase factors turned on contains a total of 6 parameters, 3 of these must be complex phase factors.

A common parametrization of $U$ in terms of mixing angles and phases is \cite{r16}
\beq
\begin{array}{l}
{\hskip -.5cm} U  =   \left[ \begin{array}{ccc}
1 & 0 & 0 \\ 0 & \phantom{-}c_{23} & s_{23} \\ 0 & -s_{23} & c_{23}
		\end{array}  \right]
		\left[ \begin{array}{ccc}
c_{13} & 0 & s_{13}e^{-i\delta} \\ 0 & 1 & 0 \\ -s_{13}e^{i\delta} &  0 & c_{13}
		\end{array}  \right]
		\left[ \begin{array}{ccc}
\phantom{-}c_{12} & s_{12}  & 0 \\ -s_{12} & c_{12}  & 0 \\  0 & 0 & 1
		\end{array}  \right]  \;   \times  \nonumber \\
 {\hskip 5cm} \times\;	\left[ \begin{array}{ccc} 
e^{i\frac{\alpha_1}{2}} & 0 & 0 \\ 0 & e^{i\frac{\alpha_2}{2}} & 0 \\ 0 & 0 & 1
		\end{array}  \right]   \nonumber \\  \vbox{\vskip 1cm}
{\hskip -1.15cm} =   \left[  \begin{array}{ccc}
c_{12}c_{13}e^{i\frac{\alpha_1}{2}} & s_{12}c_{13}e^{i\frac{\alpha_2}{2}} & s_{13}e^{-i\delta} \\
(-s_{12}c_{23} - c_{12}s_{23}s_{13}e^{i\delta})e^{i\frac{\alpha_1}{2}} & 
( \phantom{-}c_{12}c_{23} - s_{12}s_{23}s_{13}e^{i\delta}) e^{i\frac{\alpha_2}{2}} & s_{23}c_{13}  \\
( \phantom{-}s_{12}s_{23} - c_{12}c_{23}s_{13}e^{i\delta}) e^{i\frac{\alpha_1}{2}} & 
(-c_{12}s_{23} - s_{12}c_{23}s_{13}e^{i\delta})e^{i\frac{\alpha_2}{2}} & 
c_{23}c_{13}
		\end{array}  \right]  .
\end{array}
\label{eq38}
\eeq
Here, $c_{ij} \equiv \cos \theta_{ij}$ and $s_{ij} \equiv \sin \theta_{ij}$, where $\theta_{12}, \; \theta_{13}$, and $\theta_{23}$ are the three mixing angles, and $\delta, \; \alpha_1$, and $\alpha_2$ are the three phases. The phase $\delta$, referred to as a Dirac phase, is the leptonic analogue of the single phase that may be found in the 3x3 quark mixing matrix. 
The phases $\alpha_1$ and $\alpha_2$, known as Majorana phases, are the extra physically-significant phases that $U$ may contain when the neutrino mass eigenstates are Majorana particles. As may be seen in \Eq{38}, the phase $\alpha_1$ is common to all elements of the first column of $U$. Thus, it could be removed from $U$ by phase-redefining the neutrino $\nul{1}$. 
Similarly, $\alpha_2$ could be removed by redefining $\nul{2}$. However, as we have mentioned, when neutrinos are Majorana particles, phases removed from $U$ by phase-redefining neutrinos simply reappear elsewhere, and continue to have the same physical consequences they had when located in $U$ \cite{r7}.

At the origin of coordinates, $x^\mu = 0$, the weak interaction ${\cal L}_W$ of \Eq{34} transforms under CP as
\beq
(\mbox{CP}) {\cal L}_W (\mbox{CP})^{-1} =  -\frac{g}{\sqrt{2}} W_\rho^- \overline{\ell_L} \gamma^\rho U^* \nul{L} - \frac{g}{\sqrt{2}} W_\rho^+\overline{\nul{L}} \gamma^\rho U^T \ell_L ~~ .
\label{eq39}
\eeq
In writing this expression, we have taken arbitrary phase factors that in principle could be present to be unity. Comparing the CP-mirror-image of ${\cal L}_W$ in \Eq{39} with ${\cal L}_W$ itself, \Eq{34}, we see that if $U^* \neq U\,$---that is, if $U$ contains some of the phases $\delta, \; \alpha_1$, and $\alpha_2$, so that it is not real---then the weak interaction is not CP invariant. In our discussions of neutrino oscillation and double beta decay, we will see examples of CP-violating physical effects that these phases can produce.

\section{What is a Majorana neutrino?}
\label{s2}

As we have seen, the see-saw mechanism predicts that neutrinos are Majorana particles. We have also seen that, quite apart from the specific details of the see-saw mechanism, it is rather likely that nature contains Majorana neutrino mass terms. 
From the procedure we followed to diagonalize the combined Majorana and Dirac mass term of \Eq{7a} [cf. Eqs.~(\ref{eq7a})-(\ref{eq15}) and accompanying discussion], it is clear that when Majorana mass terms are present, the neutrino mass eigenstates are Majorana particles. 
Thus, it is rather likely that neutrinos are indeed Majorana particles. Since the behavior of Majorana neutrinos can---at first---be a bit puzzling, it is worth trying to clarify the nature of these particles.

A Majorana neutrino mass eigenstate $\nu_i$ is a particle whose field goes into itself under charge conjugation. Thus, the neutrino is identical to its antiparticle: $\nu_i = \bar{\nu_i}$. Now, in descriptions of neutrino processes, it is sometimes {\em assumed} that there is a conserved lepton number $L$, with $L$ (negatively-charged lepton) = $L$ (neutrino) = $-L$ (positively-charged lepton) = $-L$ (antineutrino) = 1. 
Particles are then identified as neutrinos or antineutrinos in accordance with the process through which they are produced. For example, if the production process is $\pi^+ \rightarrow \mu^+ + \nu_\mu$, the outgoing neutral particle is identified as a neutrino, not an antineutrino, because $L(\pi^+) = 0$ and $L(\mu^+) = -1$, and it is being assumed that $L$ is conserved. Similarly, if the production process is $\pi^- \rightarrow \mu^- + \overline{\nu_\mu}$, the outgoing neutral particle is identified as an antineutrino. 
Now, we know that, when interacting in a detector, the ``neutrino'' produced in $\pi^+$ decay will create a $\mu^-$, while the ``antineutrino'' produced in $\pi^-$ decay will create a $\mu^+$. (For simplicity, we are disregarding mixing and neutrino oscillation.) This behavior appears to suggest that ``neutrinos'' and ``antineutrinos'' are different particles, and that $L$ is indeed conserved. But there is another, equally viable, interpretation of this behavior. 
We know from measurements of the muon polarization in pion decays that the ``$\nu_\mu$'' produced in $\pi^+ \rightarrow \mu^+ + \nu_\mu$ has LH (negative) helicity, while the ``$\overline{\nu_\mu}$'' produced in $\pi^- \rightarrow \mu^- + \overline{\nu_\mu}$ has RH (positive) helicity. Let us now assume that nature contains Majorana mass terms, so that lepton number $L$ is not conserved, and neutrinos are Majorana particles. 
For simplicity, we also continue to neglect mixing, so that $\nu_\mu$ is a mass eigenstate. Then, {\em for a given helicity} $h, \; \nu_\mu$ and $\overline{\nu_\mu}$ are the same particle. Nevertheless, the neutral particles produced in $\pi^+$ and $\pi^-$ decay still differ from each other, because they have opposite helicity. Under the assumption that they are Majorana neutrinos, helicity is the {\em only} difference between them. 
But helicity is a sufficient difference to explain why the neutral particle coming from $\pi^+$ decay will yield a $\mu^-$ when it interacts, while the one coming from $\pi^-$ decay will yield a $\mu^+$. After all, the weak interaction is maximally parity violating, so it is not surprising at all that oppositely polarized particles interact differently. 
Indeed, it is easily verified that in the charged current weak interaction of \Eq{34}, the first term completely dominates for an incoming Majorana neutrino with negative helicity, but the second one completely dominates for the same incoming neutrino when its helicity is positive. As we see, the first term will create a negatively charged lepton, but the second term will create a positively charged one. Thus, what happens when a neutrino interacts can be understood without invoking a conserved lepton number. 
It can be explained by assuming that neutrinos are Majorana particles, and simply noting that by reversing the helicity of a Majorana neutrino, we can reverse the charge of the lepton this neutrino creates when it interacts. In this picture, the role played by the ``neutrino'' when $L$ conservation is assumed is played by the LH helicity state of the Majorana neutrino, and the role played by the ``antineutrino'' is played by the RH helicity state of the same particle.

\begin{figure}[hbt]
\begin{center}
\includegraphics[width=11cm]{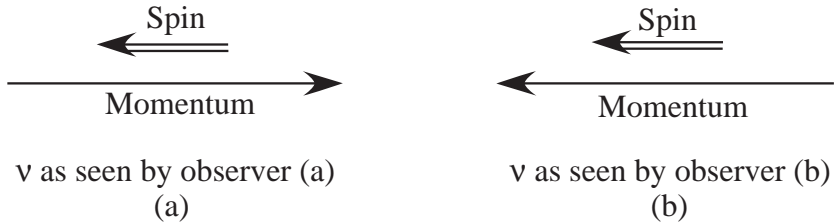}
\end{center}
\caption{A neutrino as seen from two different frames of reference.}
\label{f2-1}
\end{figure}

Correlating the charge of a produced lepton with the helicity of the Majorana neutrino that produces it leads to a puzzle. Suppose a Majorana neutrino, as seen by observer (a), is moving to the right with LH helicity, as shown in Fig.~\ref{f2-1}(a). As seen by observer (b), who is moving to the right faster than the neutrino, the latter is moving to the left. However, its spin is still pointing to the left, just as it was for observer (a). Thus, as seen by observer (b), the neutrino has {\em RH} helicity, as shown in Fig.~\ref{f2-1}(b). 
Now, suppose our neutrino interacts with a target that is at rest in the frame of observer (a) [frame (a)], and creates a $\mu^-$, a lepton with the charge expected in view of the neutrino's helicity. But, as seen by observer (b), this same neutrino has {\em RH} helicity. Does this mean that, as seen from the frame of observer (b) [frame (b)], the neutrino's interaction with the target produces a $\mu^+$, rather than a $\mu^-$? Clearly (we hope!), it had better not mean that. Lorentz transforming the $\mu^-$ from frame (a) to frame (b) certainly does not change its electric charge. 
Consequently, the neutrino collision with the target at rest in frame (a) yields a $\mu^-$ regardless of whether the collision is viewed from frame (a) or frame (b). But how can this be the case, given that, when an incident neutrino has RH helicity, as ours does in frame (b), the weak interaction of \Eq{34} appears to strongly favor the production of a positively charged lepton over that of a negatively-charged one?

The solution to this puzzle is that any collision between a neutrino and a target depends on {\em two} weak currents: the leptonic current in \Eq{34}, {\em and} a current for the target. Each of these currents is a Lorentz four-vector, and can look very different in different frames. But the amplitude for the collision is the scalar product of the two currents, and a scalar product of two four-vectors is Lorentz invariant. Thus, the result of the collision, and in particular the charge of the produced lepton, will be the same as seen by all observers.

\begin{figure}[bht]
\begin{center}
\includegraphics[width=11cm]{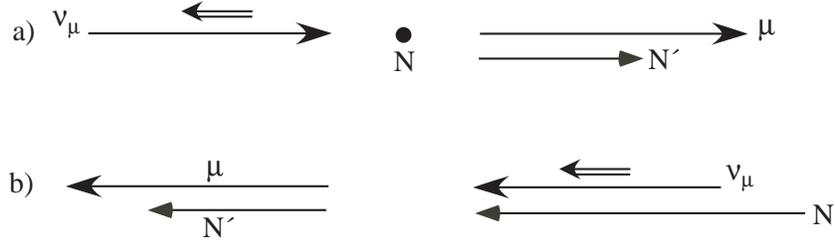}
\end{center}
\caption{a) The forward reaction $\nu_\mu + N \rightarrow \mu + N^\prime$ as seen in the $N$ rest frame, where the $\nu_\mu$ has LH helicity. Single lines show momenta, and the double line shows the $\nu_\mu$ spin. b) The same reaction as seen in a frame in which all particles move at highly relativistic speeds in a direction opposite to that of $\nu_\mu$ in the N rest frame.}
\label{f2-2}
\end{figure}

To illustrate this point, let us consider the collision between a Majorana muon neutrino $\nu_\mu$ and a spinless target $N$ (a spinless nucleus, for example). We neglect mixing, so that $\nu_\mu$ is a mass eigenstate. We assume that $\nu_\mu$ has LH helicity in the rest frame of $N$, and that the collision produces an outgoing muon and a spinless nuclear recoil $N^\prime$. Given the $\nu_\mu$ helicity, we expect the probability for the muon to be negative to far outweigh that for it to be positive, 
but we allow for both possibilities, and compute the amplitudes for the two reactions $\nu_\mu + N \rightarrow \mu^\mp + N^\prime_\pm$, where $N^\prime_+(N^\prime_-)$ is a nuclear recoil whose charge is one unit greater (less) that that of $N$. To keep the illustrative calculation simple, we take the matrix element of the nuclear target weak current, $J_N^\rho$, to have the form
\beq
\langle N^\prime_\pm(k^\prime) | J_N^\rho | N(k) \rangle = c (k+k^\prime)^\rho ~~ .
\label{eq2.1}
\eeq
Here, $k$ and $k^\prime$ are, respectively, the four-momenta of $N$ and $N^\prime_\pm$, and $c$ is a constant whose value we assume to be the same for $N^\prime_+$ and $N^\prime_-$. We consider the case of forward scattering, in which, in the $N$ rest frame, the $\mu$ and the $N^\prime$ both leave the collision with momenta parallel to the momentum of the incident $\nu_\mu$. The reaction as seen in this frame is depicted in Fig.~\ref{f2-2}(a). 
We assume that in this frame all particles, save the initial nucleus, are highly relativistic, and that, to a sufficiently good approximation, the initial nucleus and the nuclear recoil have the same mass. For this case, we find by explicit calculation in the $N$ rest frame, using the leptonic current of \Eq{34} and the nuclear one of \Eq{2.1}, that
\beq
\frac{A[\nu_\mu\mbox{(LH)} + N \rightarrow \mu^+ + N^\prime_-]}
     {A[\nu_\mu\mbox{(LH)} + N \rightarrow \mu^- + N^\prime_+]}
  \cong \frac{1}{2} \frac{m_\nu}{E_\nu} \frac{m_\mu}{E_\mu} \frac{E_{N^\prime}}{m_{N^\prime}}~~ .
\label{eq2.2}
\eeq
Here, $A[\ldots]$ is the amplitude for the process in the bracket, and $E_\nu,\; E_\mu$, and $E_{N^\prime}$ are, respectively, the energies of the neutrino, muon, and nuclear recoil, whose masses are, respectively, $m_\nu,\; m_\mu$, and $m_{N^\prime}$. As expected, $\mu^-$ production dominates over $\mu^+$ production because of the small value of $m_\nu / E_\nu$, and in the limit that $m_\nu / E_\nu \rightarrow 0$, this dominance is total.

Next, we calculate the amplitudes for $\mu^+ N^\prime_-$ and $\mu^- N^\prime_+$ production in a frame where all particles are highly relativistic, and all of them, including the $\nu_\mu$ and $N$, move in the direction opposite to that of $\nu_\mu$ in the $N$ rest frame. The view from this frame, in which the $\nu_\mu$ has RH helicity, is shown in Fig.~\ref{f2-2}(b). By explicit calculation in this frame, we find that
\begin{eqnarray} 
\frac{A[\nu_\mu\mbox{(RH)} + N \rightarrow \mu^+ + N^\prime_-]}
     {A[\nu_\mu\mbox{(RH}) + N \rightarrow \mu^- + N^\prime_+]}  &  =  &
\nonumber \\  
& \hbox{\hskip -8.6cm} = &  \hbox{\hskip -4.5cm}
\frac {E^*_\nu+m_\nu+|\vec{p}^*_\nu|}{E^*_\nu+m_\nu-|\vec{p}^*_\nu|}\;
\frac {E^*_\mu+m_\mu+|\vec{p}^*_\mu|}{E^*_\mu+m_\mu-|\vec{p}^*_\mu|}\;
\frac{(E^*_N+E^*_{N^\prime})-(|\vec{k}^*|+|\vec{k^\prime}^*|)}{(E^*_N+ E^*_{N^\prime})+(|\vec{k}^*|+|\vec{k^\prime}^*|)} ~ . 
\label{eq2.3}
\end{eqnarray}
Here, $A$ once again denotes an amplitude, and the listed $\nu_\mu$ helicity is the one seen in the new frame. The quantities $E^*_\nu$ and $|\vec{p}^*_\nu|$ are, respectively, the energy and momentum of the neutrino in this frame, and similarly for $E^*_\mu$ and $|\vec{p}^*_\mu|,\; E^*_N$ and $|\vec{k}^*|$, and $E^*_{N^\prime}$ and $|\vec{k^\prime}^*|$.

It is tedious, but straightforward, to re-express the right-hand side of \Eq{2.3} in terms of quantities in the $N$ rest frame. When one does this, one finds that the ratio of amplitudes in \Eq{2.3} is exactly the same as the ratio of amplitudes in \Eq{2.2}. That is, the relative rates at which $\mu^+ N^\prime_-$ and $\mu^- N^\prime_+$ are produced are exactly the same in both of the frames we have considered, as demanded by Lorentz invariance. 
In particular, $\mu^-$ production dominates over $\mu^+$ production in both frames, despite the fact that in one of the frames the incoming neutral lepton is right-handed.

\section{Neutrino flavor change}
\label{s3}

There is now a strong conviction that neutrinos do have nonzero masses and mix. As indicated at the start of this chapter, this conviction is based on the compelling evidence that neutrinos can change from one flavor to another. In this section, we shall briefly review the physics of neutrino flavor change, and see why this phenomenon implies neutrino masses and mixing. 

\begin{figure}[hbt]
\begin{center}
\includegraphics[width=11cm]{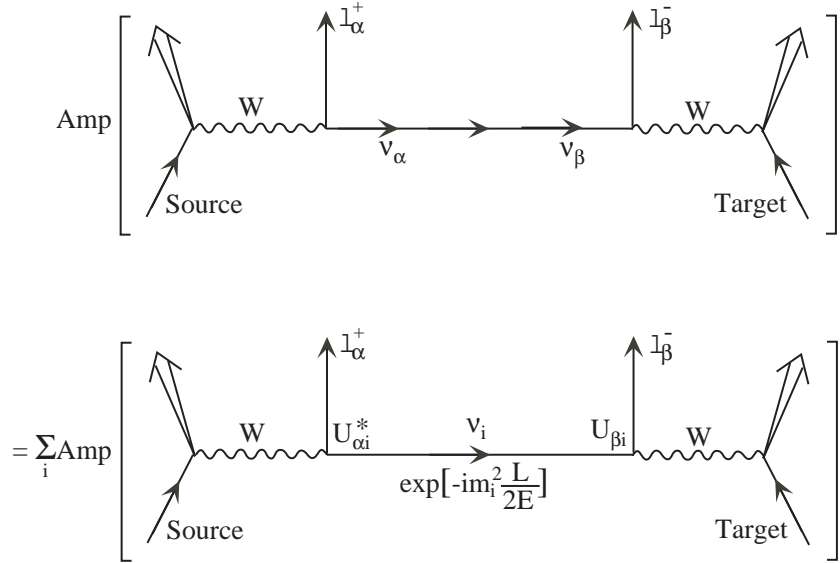}
\end{center}
\caption{Neutrino flavor change {\it in vacuo}. As shown in the upper diagram, the neutrino is created together with a charged lepton $\ell_\alpha^+$ by a Source. After traveling a distance $L$, it interacts with a target and produces a second charged lepton $\ell_\beta^-$. As shown in the lower diagram, the amplitude for this process is a sum over the contributions of all the neutrino mass eigenstates $\nu_i$.}
\label{fig3-1}
\end{figure}

Neutrino flavor change {\it in vacuo} is the process in which a neutrino is created together with a charged lepton $\ell_\alpha$ of flavor $\alpha$, then travels a macroscopic distance $L$ in vacuum, and finally interacts with a target to produce a second charged lepton $\ell_\beta$ whose flavor $\beta$ is different from that of the first charged lepton. 
That is, in the course of traveling from source to target, the neutrino morphs from a $\nu_\alpha$ to a $\nu_\beta$. 
The process, commonly referred to as $\nu_\alpha \rightarrow \nu_\beta$ oscillation, is depicted in the upper diagram of Fig.~\ref{fig3-1}. 
As shown in the lower diagram of Fig.~\ref{fig3-1}, the intermediate neutrino can be any of the (light) mass eigenstates $\nu_i$, and the amplitude for the oscillation is the coherent sum of the contributions of the various mass eigenstates. (From this point on, we use the simplified notation $\nu_i$, without a superscript ``Light'', to mean a light neutrino mass eigenstate.) 
The contribution of a given $\nu_i$ is a product of three factors: First, from \Eq{1} or (\ref{eq37}), the amplitude for the created $\nu_\alpha$ to be the mass eigenstate $\nu_i$ is $U^*_{\alpha i}$. Secondly, the amplitude for this $\nu_i$ to travel a distance $L$ if the neutrino energy is E is $\exp[-im_i^2 L/2E]$ \cite{r17}. 
Finally, the amplitude for $\nu_i$, having arrived at the target, to produce the particular charged lepton $\ell_\beta^-$ is, from \Eq{37}, $U_{\beta i}$. Thus, the amplitude Amp$[\nu_\alpha \rightarrow \nu_\beta]$ for 
$\nu_\alpha \rightarrow \nu_\beta$ oscillation is given by
\beq
\mbox{Amp}[\nu_\alpha \rightarrow \nu_\beta] = \sum_i U^*_{\alpha i} e^{-im_i^2 \frac{L}{2E}} U_{\beta i} ~~ ,
\label{eq3.1}
\eeq
where the sum runs over all the light mass eigenstates. Squaring this relation and using the (at least approximate) unitarity of the mixnig matrix $U$, we find that the probability $P(\nu_\alpha \rightarrow \nu_\beta)$ for $\nu_\alpha \rightarrow \nu_\beta$ oscillation is given by \cite{r17}
\begin{eqnarray}
P(\nu_\alpha\rightarrow\nu_{\beta})  & = & 
|\mbox{Amp}[\nu_\alpha \rightarrow \nu_\beta]|^2 =\delta_{\alpha\beta} 
\nonumber  \\
& & \mbox{} - 4\sum_{i>j} \Re\,(U^*_{\alpha i} U_{\beta i} U_{\alpha j} U^*_{\beta j})
\sin^2[\Delta m^2_{ij}(L/4E)]   \nonumber  \\
& & \mbox{} + 2\sum_{i>j} \Im \,(U^*_{\alpha i} U_{\beta i} U_{\alpha j} U^*_{\beta j}) 
\sin[\Delta m^2_{ij}(L/2E)] ~~ .
\label{eq3.2}
\end{eqnarray}
where $\Delta m^2_{ij} \equiv m^2_i - m^2_j$. This expression for $P(\nu_\alpha \rightarrow \nu_\beta)$ is valid for an arbitrary number of neutrino mass eigenstates, and holds whether $\beta$ is different from $\alpha$ or not. However, we see that if all the neutrino masses, and consequently all the splittings $\Delta m^2_{ij}$, vanish, then P$(\nu_\alpha\rightarrow\nu_{\beta}) = \delta_{\alpha\beta}$. Thus, the oscillation {\it in vacuo} of $\nu_\alpha$ into a {\em different} flavor $\nu_\beta$ implies neutrino mass. From \Eq{3.1}, we see that this change of flavor also implies neutrino mixing: In the absence of mixing, the $U$ matrix is diagonal, so that Amp[$\nu_\alpha \rightarrow \nu_\beta$] vanishes if $\beta \neq \alpha$. Finally, from \Eq{3.2} we see that the probability for neutrino oscillation really does oscillate as a function of $L/E$, justifying the name ``oscillation''.

Assuming that CPT invariance holds,
\beq
P(\overline{\nu_\alpha}\rightarrow\overline{\nu_\beta}) = 
P(\nu_\beta\rightarrow\nu_{\alpha}) ~~ .
\label{eq3.3}
\eeq
However, from \Eq{3.2} we see that
\beq
P(\nu_\beta\rightarrow\nu_{\alpha};U) = 
P(\nu_\alpha\rightarrow\nu_{\beta}; U^*) ~~ .
\label{eq3.4}
\eeq
Thus,
\beq
P(\overline{\nu_\alpha}\rightarrow\overline{\nu_\beta};U) = 
P(\nu_\alpha\rightarrow\nu_{\beta}; U^*) ~~ .
\label{eq3.5}
\eeq
That is, the probability for $\overline{\nu_\alpha} \rightarrow \overline{\nu_\beta}$ is the same as for $\nu_\alpha \rightarrow \nu_{\beta}$, except that $U$ is replaced by $U^*$. But this means that if $U$ is not real, then P($\overline{\nu_\alpha} \rightarrow \overline{\nu_\beta}$) differs from P($\nu_\alpha\rightarrow\nu_{\beta}$) by a reversal of the last term of \Eq{3.2}. This difference is a violation of CP invariance, which would require $\nu_\alpha \rightarrow \nu_{\beta}$ and $\overline{\nu_\alpha} \rightarrow \overline{\nu_\beta}$ to have equal probability.

Neutrino oscillation depends on the interference of different contributions to an amplitude [cf. \Eq{3.1}], so it is a quintessentially quantum mechanical phenomenon. It raises a number of subtle questions, some of which have been addressed  by treatments based on wave packets \cite{r18}. 
However, it has also been shown that for a number of the oscillation observations that are made in practice, a wave packet treatment is not necessary \cite{r19}. Sophisticated analyses of oscillation continue to yield new insights \cite{r20}. However, they lead to the same oscillation probability as we have obtained here.

If neutrinos pass through enough matter between their source and a target detector, then their coherent forward scattering from particles in this matter can significantly modify their oscillation pattern. This is true even if, as in the Standard Model, their forward scattering from other particles does not by itself change neutrino flavor. 
Flavor change in matter that grows out of an interplay between flavor-nonchanging neutrino-matter interactions and neutrino mass and mixing is known as the Mikheyev-Smirnov-Wolfenstein (MSW) effect \cite{r20a}.

To treat a neutrino in matter, it is convenient to describe its state by a column vector in flavor space,
\beq \left[  \begin{array}{c}
					a_e (t)  \\  a_\mu (t)  \\  a_\tau (t)   
					\end{array}  \right] ~~ ,
\label{eq3.6}
\eeq
where $a_e (t)$ is the amplitude for the neutrino to be a $\nu_e$ at time $t$, and similarly for the other flavors. The time evolution of the neutrino state is then described by a Schr\"{o}dinger equation in which the Hamiltonian ${\cal H}$ is a 3x3 matrix that acts on this column vector \cite{r21}. To illustrate, we shall make the simplifying assumption that we are dealing with an effectively ``two-neutrino'' problem, in which only $\nu_e$ and $\nu_\mu$, and two corresponding mass eigenstates $\nu_1$ and $\nu_2$, matter. Then the neutrino is described by a two-component column vector,
\beq \left[  \begin{array}{c}
					a_e (t)  \\  a_\mu (t)
					\end{array}  \right] ~~ ,
\label{eq3.7}
\eeq
and ${\cal H}$ is 2x2. If our neutrino is traveling {\it in vacuo}, then mixing is described by the vacuum mixing matrix
\begin{eqnarray}
& &\hbox{\hskip 1.2cm} \begin{array}{cc} 1 \hbox{\hskip 1.1cm} & 2\end{array} \nonumber \\
U_V &  =  & \begin{array}{c} e \\ \mu \end{array}
					\left[  \begin{array}{cc}
					\phantom{-}\cos \theta_V  &  \sin \theta_V  \\
					-\sin\theta_V  &  \cos \theta_V
					\end{array}  \right]  ~~ ,
\label{eq3.8}
\end{eqnarray}
in which $\theta_V$ is the mixing angle {\it in vacuo}, and the symbols above and to the left of the matrix label the columns and rows.  It is easy to show that, apart from an irrelevant multiple of the identity, ${\cal H}$ {\it in vacuo} is then \cite{r21}
\beq
{\cal H}_V = \frac{\Delta m_V^2}{4E} \left[  \begin{array}{cc}
					-\cos 2\theta_V  &  \sin 2\theta_V  \\
					\phantom{-}\sin 2 \theta_V  &  \cos 2\theta_V
					\end{array}  \right]  ~~ .
\label{eq3.9}
\eeq
Here, $\Delta m_V^2 \equiv m_2^2 - m_1^2$ is the (mass)$^2$ splitting {\it in vacuo}, and $E$ is the neutrino energy. One can straightforwardly show that this ${\cal H}$$_V$ predicts that the probability $P_V(\nu_e \rightarrow \nu_\mu)$ for $\nu_e \rightarrow \nu_\mu$ oscillation {\it in vacuo} is given by
\beq
P_V(\nu_e \rightarrow \nu_\mu) = \sin^2 2\theta_V \sin^2 (\Delta m^2_V \frac{L}{4E} ) ~~ .
\label{eq3.10}
\eeq
This is the famous formula for two-neutrino oscillation {\it in vacuo}. It follows also from \Eq{3.2} for the special case of two neutrinos, if we take $\alpha = e,\; \beta = \mu, \; i = 2,\; j = 1,\; \Delta m_{21}^2 \equiv \Delta m_V^2$, and $U$ to be the matrix $U_V$ of \Eq{3.8}.

In matter, $W$-exchange-induced coherent forward scattering of $\nu_e$ from ambient electrons adds an interaction energy $V$ to the $\nu_e - \nu_e$ element of ${\cal H}$. (The $\nu_\mu - \nu_\mu$ element is not affected, because the reaction $\nu_\mu e \rightarrow \nu_\mu e$ cannot be induced by $W$ exchange.)
Obviously, $V$ must be proportional to $G_F$, the Fermi constant, and to $N_e$, the number of electrons per unit volume. Indeed, it can be shown that \cite{r22}
\beq
V = \sqrt{2}\,G_F\, N_e ~~ .
\label{eq3.11}
\eeq
In addition, $Z$-exchange-mediated scattering from ambient particles adds a further interaction energy to all diagonal elements of ${\cal H}$. However, since the $Z$ coupling to neutrinos is flavor independent, this further addition to ${\cal H}$ is a multiple of the identity matrix, and no such addition has any effect on neutrino flavor oscillation \cite{r21}. Thus, we may safely omit the $Z$-exchange-induced energy. Then the 2x2 Hamiltonian in matter is
\beq
{\cal H} = \frac{\Delta m_V^2}{4E} \left[  \begin{array}{cc}
									-\cos 2\theta_V  &  \sin 2\theta_V  \\
					\phantom{-}\sin 2 \theta_V  &  \cos 2\theta_V
					\end{array}  \right]   + 
					\left[ \begin{array}{cc} V & 0 \\ 0 & 0 
					\end{array}\right] ~~ .
\label{eq3.11a}
\eeq
Harmlessly adding to this ${\cal H}$ the multiple - V/2 of the identity, we may rewrite it as
\beq
{\cal H} = \frac{\Delta m_M^2}{4E} \left[  \begin{array}{cc}
									-\cos 2\theta_M  &  \sin 2\theta_M  \\
					\phantom{-}\sin 2 \theta_M  &  \cos 2\theta_M
					\end{array}  \right]    ~~ .
\label{eq3.12}
\eeq
Here,
\beq
\Delta m_M^2 = \Delta m_V^2 \sqrt{\sin^2 2\theta_V + (\cos 2\theta_V - x)^2}
\label{eq3.13}
\eeq
is the effective mass splitting in matter, and
\beq
\sin^2 2\theta_M = \frac{\sin^2 2\theta_V}{\sin^2 2\theta_V + 
(\cos 2\theta_V - x)^2}
\label{eq3.14}
\eeq
is the effective mixing angle in matter. In these expressions,
\beq
x \equiv \frac{V}{(\Delta m_V^2/2E)}
\label{eq3.15}
\eeq
is a dimensionless measure of the relative importance of the matter interaction on the neutrino behavior.

If a neutrino travels through matter of constant density, then ${\cal H}$, \Eq{3.12}, is a position-independent constant. As we see, it is exactly the same as the vacuum Hamiltonian, \Eq{3.9}, except that the vacuum mass splitting and mixing angle are replaced by their values in matter. As a result, the oscillation probability is given by the usual formula, \Eq{3.10}, but with the mass splitting and mixing angle replaced by their values in matter. 
The latter values can differ markedly from their vacuum counterparts. A striking example is the case where the vacuum mixing angle $\theta_V$ is very small, but $x \cong \cos 2\theta_V$. Then, as we see from \Eq{3.14}, $\sin^2 2\theta_M \cong 1$. Matter interaction has promoted a very small mixing angle into a maximal one.

One important example of neutrino propagation in matter is the journey of solar neutrinos, which are created as electron neutrinos in the center of the sun, outward through solar material. Of course, the electron density encountered by these neutrinos is not a constant, so ${\cal H}$ depends on the distance $r$ from the center of the sun. Nevertheless, under certain conditions the propagation of the neutrinos is adiabatic. 
That is, the electron density $N_e (r)$ varies slowly enough that one may solve the Schr\"{o}dinger equation for neutrino propagation for one $r$ at a time, and then patch together the solutions. This is true, in particular, for the so-called Large Mixing Angle (LMA) version of the MSW picture of what happens to the solar neutrinos, which is the most favored explanation of their observed behavior.

In the LMA MSW scenario, $\Delta m_V^2 \sim 5 \times 10^{-5}$ eV$^2$ \cite{r23}. For the most closely scrutinized solar neutrinos, the ones from $^8$B decay, typical energies $E$ are 6-7 MeV. For these neutrinos, $\Delta m_V^2 / 4E \sim 0.2 \times 10^{-5}$ eV$^2$/MeV. Now, at $r \simeq 0$, where the solar neutrinos are born, the electron density $N_e \simeq 6 \times 10^{25}$/cm$^3$ \cite{r24}. 
This value yields for the interaction energy $V$ at $r \simeq 0$, the value $V \sim  0.75 \times 10^{-5}$ eV$^2$/MeV. Consequently, where the neutrinos are born, the interaction (second) term of the Hamiltonian ${\cal H}$ of \Eq{3.11a} dominates over the vacuum (first) term, at least to some extent. As a result, ${\cal H}$ is approximately diagonal at $r \simeq 0$. 
This means that at birth, a $^8$B neutrino is not only a $\nu_e$ but also, approximately, in an eigenstate of ${\cal H}$. Since $V>0$, the neutrino is in the heavier of the two eigenstates. Then it propagates outward adiabatically. This means that it continues to be in an eigenstate of ${\cal H}$---an $r$-dependent eigenstate that changes slowly as ${\cal H}$ changes. It will then emerge from the sun as one of the two eigenstates of the zero-density (vacuum) Hamiltonian. 
That is, our neutrino leaves the sun as one of the mass eigenstates of ${\cal H}_V$. Since, as one may readily verify, the eigenlevels of ${\cal H}$, \Eq{3.11a}, never cross, and the neutrino started in the heavier eigenlevel at $r \simeq 0$, it will leave the sun as the heavier of the two mass eigenstates of ${\cal H}_V$. 
If we define $\Delta m_V^2 = m_2^2 - m_1^2$ to be positive, then this is the eigenstate called $\nu_2$. Being an eigenstate of the vacuum Hamiltonian, this state will propagate without mixing all the way to the surface of the earth. Now, from \Eq{3.8}, $\nu_2$ has the flavor composition
\beq
|\nu_2\rangle = |\nu_e\rangle \sin\theta_V + |\nu_\mu \rangle\cos\theta_V ~~ .
\label{eq3.16}
\eeq
The probability that a $^8$B solar neutrino still has the $\nu_e$ flavor with which it was born when it arrives at earth is just the $\nu_e$ fraction of this state, $\sin^2 \theta_V$.

When information from atmospheric neutrino oscillation is taken into account, one learns that the ``other flavor'' with which solar electron neutrinos mix is not $\nu_\mu$ but a 50-50 mixture of $\nu_\mu$ and $\mu_\tau$. However, if one simply understands ``$\nu_\mu$'' in our analysis of the solar neutrinos to be a shorthand for this 50-50 mixture, then that analysis remains valid.

Like oscillation {\it in vacuo}, neutrino flavor change in matter requires neutrino masses and mixing. If either $\Delta m_V^2$ or $\theta_V$ vanishes, then the Hamiltonian in matter, \Eq{3.11a}, is diagonal. Thus, a neutrino born with a given flavor will retain that flavor forever.

Flavor change has been reported for atmospheric neutrinos, solar neutrinos, and the accelerator neutrinos studied by the Liquid Scintillator Neutrino Detector (LSND) experiment. Each of these three reported flavor changes calls for a splitting $\Delta m^2$ that is of a different order of magnitude than the ones called for by the other two. Obviously, these three very different splittings cannot all be accommodated if there are only three neutrino mass eigenstates, since there are then only three splittings $\Delta m^2_{ij}$, and they obviously satisfy
\beq
\Delta m^2_{32} + \Delta m^2_{21} + \Delta m^2_{13} = 0 ~~ .
\label{eq3.17}
\eeq
Thus, if all three reported flavor changes prove to be genuine, then nature must contain at least four neutrino mass eigenstates $\nu_i$. Now, three linear combinations of these $\nu_i$, namely $\nu_e,\; \nu_\mu$, and $\nu_\tau$, couple to the $W$ boson and one of the three charged leptons. 
If there are exactly four $\nu_i$, then there is a fourth linear combination of them, $\nu_s$, orthogonal to $\nu_e,\; \nu_\mu$, and $\nu_\tau$, which has no charged-lepton partner, and hence cannot couple to the $W$. Moreover, since the decays $Z \rightarrow \nu_\alpha \bar{\nu}_\alpha$ of the $Z$ into neutrino pairs are found to produce only three distinct neutrino flavors \cite{r25}, the fourth neutrino $\nu_s$ evidently does not couple to the $Z$ either. 
Thus, $\nu_s$ does not have any of the Standard Model weak couplings. Such a neutrino is called ``sterile''. Obviously, it is quite unlike the ``active'' neutrinos, $\nu_e,\; \nu_\mu$, and $\nu_\tau$. Consequently, it will be very interesting to see whether all three of the reported neutrino flavor changes are confirmed, so that nature must contain a sterile neutrino.

\section{Double beta decay}
\label{s4}

Given the theoretical expectation that neutrinos are Majorana particles, it would obviously be desirable to confirm experimentally that this is indeed the case. As mentioned in Sec.~\ref{s1}, the observation of neutrinoless double beta decay, the reaction Nucl $\rightarrow$ Nucl$^\prime + 2e^-$, would provide the sought-for confirmation \cite{r11}.

\begin{figure}[hbt]
\begin{center}
\includegraphics[height=4cm]{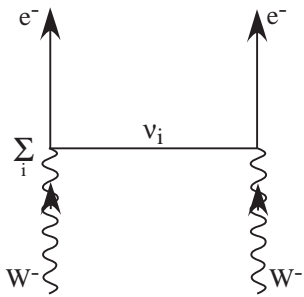}
\end{center}
\caption{The process at the heart of neutrinoless double beta decay. The exchanged particle can be any of the light neutrinos $\nu_i$.}
\label{fig4-1}
\end{figure}

If neutrinoless double beta decay (often referred to as $0\nu\beta\beta$) does occur, it is quite likely dominated by a mechanism in which the parent nucleus emits a pair of virtual $W^-$ bosons, turning into the daughter nucleus, and then the $W^-$ bosons exchange one or another of the light neutrino mass eigenstates $\nu_i$ to create the outgoing electrons. The heart of this mechanism is the second step, $W^- W^- \rightarrow e^-e^-$ via Majorana neutrino exchange. The diagram for this step is shown in Fig.~\ref{fig4-1}.
There, the Standard Model weak interaction is assumed to act at each vertex. When neutrinos and antineutrinos differ, this interaction creates the exchanged particle as an antineutrino, but can absorb it only as a neutrino. Thus, the diagram is forbidden unless neutrinos and antineutrinos do {\em not} differ---the Majorana case.

As indicated in  Fig.~\ref{fig4-1}, the amplitude for $W^- W^- \rightarrow e^-e^-$ is the coherent sum of the contributions of all the light neutrino mass eigenstates $\nu_i$. From \Eq{37}, the contribution of $\nu_i$ involves the current $\overline{\ell_{Le}} \gamma^\rho U_{ei} \nu_{Li}$, acting at both vertices. Thus, this contribution is proportional to $U_{ei}^2$. It is also proportional to $m_i$. The latter factor may be understood by recalling that the exchanged $\nu_i$ is produced as an ``antineutrino'', which in the Majorana case simply means that it has the helicity normally associated with an antineutrino. That is, it is right-handed, except for a small left-handed piece with amplitude of order $m_i /E,\; E$ being its energy. It is only this left-handed piece that the LH weak current acting to absorb the $\nu_i$ can accommodate without further suppression. Thus, the amplitude for $0\nu\beta\beta$ is proportional to a factor $m_{\beta\beta}$ given by
\beq
m_{\beta\beta} =  | \;\, \sum_i m_i U_{ei}^2 \,| ~~,
\label{eq4.1}
\eeq
and referred to as the effective neutrino mass for double beta decay.

While neutrino oscillation has provided us with the evidence that neutrino masses are nonzero, this process cannot determine the masses $m_i$ of the individual neutrino mass eigenstates. Rather, oscillation can only determine the (mass)$^2$ {\em splittings} $\Delta m^2_{ij}$, as \Eq{3.2} for $P(\nu_\alpha \rightarrow \nu_\beta)$ makes very evident. One approach to gaining some information about the $m_i$, and thereby some knowledge of the absolute scale of neutrino mass, is to look for kinematical effects of neutrino mass in the leptonic tritium decays, $^3$H $\rightarrow ^3$He $+ e^- + \bar{\nu}_i$ \cite{r26}. Another approach is to look for $0\nu\beta\beta$, since a knowledge of $m_{\beta\beta}$, \Eq{4.1}, would clearly provide at least some information on the scale of the masses $m_i$.

The effective mass $m_{\beta\beta}$ could, in principle, also provide some information on the CP-violating phases in the $U$ matrix of \Eq{38}. From \Eq{3.1}, we see that only the Dirac phase $\delta$ in this matrix can influence neutrino oscillation. Any Majorana phase, such as $\alpha_1$, is common to an entire column of $U$. 
Thus, this phase cancels out of the oscillation amplitude, in which the $\nu_i$ contribution, as we see in \Eq{3.1}, is proportional to $U^*_{\alpha i} U_{\beta i}$. On the other hand, a Majorana phase, say in the $i$'th column of $U$, would not cancel out of $m_{\beta\beta}$, since, as \Eq{4.1} shows, the $\nu_i$ contribution to $m_{\beta\beta}$ is proportional to $U^2_{ei}$, rather than $U^*_{ei}U_{ei}$. 
Thus, if we know the masses $m_i$ and the mixing angles in $U$, and we also know $m_{\beta\beta}$ with sufficient precision, we can in principle learn something about the Majorana phases, or at least demonstrate that they are present. Whether this would be feasible in practice is being explored \cite{r27}.

\section{Conclusion}
\label{Conclusion}

Neutrino flavor change, either {\it in vacuo} or in matter, implies neutrino mass and mixing. Thus, the very strong evidence for flavor change makes a compelling case that neutrinos have nonzero masses. Owing to the possibility---unique to neutrinos---of Majorana mass terms, the physics underlying neutrino mass may be quite different from that underlying the masses of the quarks and charged leptons. In addition, if Majorana mass terms are present, the neutrinos are Majorana particles, making them quite different from the other fundamental fermions.

Progress in understanding the world of neutrinos has been quite striking in recent years. However, we are still only beginning to uncover the secrets of this world. Exciting years lie ahead.

\section*{Acknowledgments}

It is a pleasure to thank Adam Para for asking the question that led us to examine explicitly how a collision between a Majorana neutrino and a target is described in different Lorentz frames. We are grateful to Susan Kayser for her very generous assistance in the production of this manuscript.

\input{reference-Boris}

\end{document}

%% file: reference-Boris.tex

%% file: chapternu.bbl
\begin{thebibliography}{99.}
\bibitem{r2} Increasingly, $U$ is referred to as the MNS matrix, or as the MNSP matrix, to honor pioneering contributions of Maki, Nakagawa, Sakata, and Pontecorvo.
\bibitem{r4} A chirally left-handed (right-handed) field $\psi_{L(R)}$ is one that obeys the relations $P_{L(R)} \psi_{L(R)} = \psi_{L(R)}$ and $P_{R(L)} \psi_{L(R)} = 0$. Here, $P_L = \frac{1}{2} (1 - \gamma_5)$[$P_R = \frac{1}{2} (1 + \gamma_5)$] is the lefthanded [right-handed] chiral projection operator. When the quanta of the field are massless, $\psi_{L(R)}$ annihilates particles of left-handed (right-handed) helicity, and creates antiparticles of right-handed (left-handed) helicity. When the quanta are not massless, there are corrections of order mass/energy to these rules.
\bibitem{r5} The author first heard the argument that follows from Belen Gavela.
\bibitem{r6} Any negative eigenvalue can be converted into a positive one by a suitable modification of $Z$.
\bibitem{r12} For simplicity, we shall not include the further very interesting possibility that there are light sterile neutrinos in addition to the three light active ones.
\bibitem{r17} B. Kayser: cf. Ref.~\cite{r16}. Some of the treatment in the present Section follows that in this reference.
\end{thebibliography}
